\documentclass[12pt]{article}

\usepackage{amsmath,amssymb,amsfonts}
\usepackage{psfrag}
\usepackage{color}
\definecolor{darkblue}{rgb}{0.1,0.1,.7}
\usepackage[colorlinks, linkcolor=darkblue, citecolor=darkblue, urlcolor=darkblue, linktocpage]{hyperref} 
\usepackage[square, comma, compress,numbers]{natbib}
\usepackage[]{amsmath}
\usepackage[]{graphicx}
\usepackage[]{latexsym}
\usepackage{geometry}
\usepackage{amscd}
\usepackage[all,cmtip]{xy}
\usepackage{mathrsfs}
\usepackage{bbold}
\usepackage{subfigure}
\usepackage[margin=10pt,font=small,labelfont=bf]{caption}
\usepackage{dsdshorthand}
\usepackage{simplewick}
\usepackage{changepage}
\numberwithin{equation}{section}

\renewcommand{\be}{\begin{eqnarray}}
\renewcommand{\ee}{\end{eqnarray}}
\newcommand{\bea}{\begin{eqnarray}}
\newcommand{\eea}{\end{eqvnarray}}

\def\beq{\begin{equation}} 
\def\eeq{\end{equation}} 
\def\<{\langle}
\def\>{\rangle}

\def\nn{\nonumber} 
 
\def\cO {{\cal O}}

\def\cN {{\cal N}}

\begin{document}

\vspace*{-.6in} \thispagestyle{empty}
\begin{flushright}
CERN-TH-2016-050
\end{flushright}
\vspace{.2in} {\Large
\begin{center}
{\bf Precision Islands in the Ising and $O(N)$ Models}
\end{center}
}
\vspace{.2in}
\begin{center}
{\bf 
Filip Kos$^{a}$, 
David Poland$^{a,b}$,
David Simmons-Duffin$^{b}$,
Alessandro Vichi$^{c}$} 
\\
\vspace{.2in} 
$^a$ {\it  Department of Physics, Yale University, New Haven, CT 06520}\\
$^b$ {\it School of Natural Sciences, Institute for Advanced Study, Princeton, New Jersey 08540}\\
$^c$ {\it Theory Division, CERN, Geneva, Switzerland}
\end{center}

\vspace{.2in}

\begin{abstract}
We make precise determinations of the leading scaling dimensions and operator product expansion (OPE) coefficients in the 3d Ising, $O(2)$, and $O(3)$ models from the conformal bootstrap with mixed correlators. We improve on previous studies by scanning over possible relative values of the leading OPE coefficients, which incorporates the physical information that there is only a single operator at a given scaling dimension. The scaling dimensions and OPE coefficients obtained for the 3d Ising model, $(\De_{\s}, \De_{\e},\lambda_{\s\s\e}, \lambda_{\e\e\e}) = (0.5181489(10), 1.412625(10), 1.0518537(41), 1.532435(19))$, give the most precise determinations of these quantities to date.
\end{abstract}

\newpage

\tableofcontents

\newpage

%%%%%%%%%%%%%%%%%%%%%%%%%%%%%%%%%%%%%%%%%%%%
%%%%%%%%%%%%                                                              %%%%%%%%%%%%%%%
%%%%%%%%%%%%     BEGIN DOCUMENT                     %%%%%%%%%%%%%%%
%%%%%%%%%%%%                                                              %%%%%%%%%%%%%%%
%%%%%%%%%%%%%%%%%%%%%%%%%%%%%%%%%%%%%%%%%%%%

\section{Introduction}
\label{sec:introduction}

The conformal bootstrap~\cite{Ferrara:1973yt,Polyakov:1974gs} in $d>2$ has recently seen an explosion of exciting and nontrivial results, opening the door to the possibility of a precise numerical classification of non-perturbative conformal field theories (CFTs) with a small number of relevant operators. Such a classification would lead to a revolution in our understanding of quantum field theory, with direct relevance to critical phenomena in statistical and condensed matter systems, proposals for physics beyond the standard model, and quantum gravity. 

One of the most striking successes has been in its application to the 3d Ising model, initiated in~\cite{ElShowk:2012ht,El-Showk:2014dwa}. In~\cite{Kos:2014bka} we found that the conformal bootstrap applied to a system of correlators $\{\<\s\s\s\s\>, \<\s\s\e\e\>, \<\e\e\e\e\>\}$ containing the leading $\Z_2$-odd scalar $\sigma$ and leading $\Z_2$-even scalar $\epsilon$ led to a small isolated allowed region for the scaling dimensions $(\De_{\s}, \De_{\e})$. In~\cite{Simmons-Duffin:2015qma} this approach was pushed farther using the semidefinite program solver \texttt{SDPB}, leading to extremely precise determinations of the scaling dimensions and associated critical exponents.\footnote{A complementary approach to solving the 3d Ising model with the conformal bootstrap was also developed in~\cite{Gliozzi:2013ysa,Gliozzi:2014jsa}.} 

In~\cite{Kos:2015mba} we found that this approach could also be extended to obtain rigorous isolated regions for the whole sequence of 3d $O(N)$ vector models, building on the earlier results of~\cite{Kos:2013tga,Nakayama:2014yia}. While the resulting ``$O(N)$ archipelago" is not yet as precise as in the case of the Ising model, it serves as a concrete example of how the bootstrap can lead to a numerical classification -- if we can isolate every CFT in this manner and make the islands sufficiently small, then we have a precise and predictive framework for understanding the space of non-perturbative conformal fixed points. If the methods can be made more efficient, it is clear that this approach may lead to solutions of longstanding problems such as determining the conformal windows of 3d QED and 4d QCD.

Compared to previous mixed-correlator studies~\cite{Kos:2014bka, Simmons-Duffin:2015qma, Kos:2015mba} (see also~\cite{Lemos:2015awa,Behan:2016dtz,Nakayama:2016jhq}), the novelty of the present work is the idea of disallowing degeneracies in the CFT spectrum by making exclusion plots in the space of OPE coefficients and dimensions simultaneously. For example, in the 3d Ising model, by scanning over possible values of the ratio $\lambda_{\e\e\e}/\lambda_{\s\s\e}$, we can impose that there is a unique $\e$ operator. This leads to a three-dimensional island in $(\De_{\s}, \De_{\e}, \lambda_{\e\e\e}/\lambda_{\s\s\e})$ space whose projection to the $(\De_{\s}, \De_{\e})$ plane is much smaller than the island obtained without doing the scan. For each point in this island, we also bound the OPE coefficient magnitude $\lambda_{\s\s\e}$. The result is a new determination of the leading scaling dimensions $(\De_{\s}, \De_{\e}) = (0.5181489(10), 1.412625(10))$, shown in figure~\ref{fig:IsingIsland}, as well as precise determinations of the leading OPE coefficients $(\lambda_{\s\s\e}, \lambda_{\e\e\e}) = (1.0518537(41), 1.532435(19))$. These scaling dimensions translate to the critical exponents $(\eta,\nu) = (0.0362978(20), 0.629971(4))$.

We repeat this procedure for 3d CFTs with $O(2)$ and $O(3)$ global symmetry, focusing on the bootstrap constraints from the correlators $\{\<\f\f\f\f\>, \<\f\f s s\>, \<ssss\>\}$ containing the leading vector $\phi_i$ and singlet $s$. We again find that scanning over the ratio of OPE coefficients $\lambda_{sss}/\lambda_{\f\f s}$ leads to a reduction in the size of the islands corresponding to the $O(2)$ and $O(3)$ vector models. The results are summarized in figure~\ref{fig:ONArchipelago}. In studying the $O(2)$ model, we are partially motivated by the present $\sim8\sigma$ discrepancy between measurements of the heat-capacity critical exponent $\alpha$ in ${}^4$He performed aboard the space shuttle STS-52~\cite{Lipa:2003zz} and the precise Monte Carlo simulations performed in~\cite{Campostrini:2006ms}. While our new $O(2)$ island is not quite small enough to resolve this issue definitively, our results have some tension with the reported ${}^4$He measurement and currently favor the Monte Carlo determinations.

This paper is organized as follows. In section~\ref{sec:crossing} we review the bootstrap equations relevant for the 3d Ising and $O(N)$ vector models and explain the scan over relative OPE coefficients employed in this work. In section~\ref{sec:results} we describe our results, and in section~\ref{sec:conclusions} we give a brief discussion. Details of our numerical implementation are given in appendix~\ref{sec:appA}.

\begin{center}
\begin{figure}[t!]
\includegraphics[width=\textwidth]{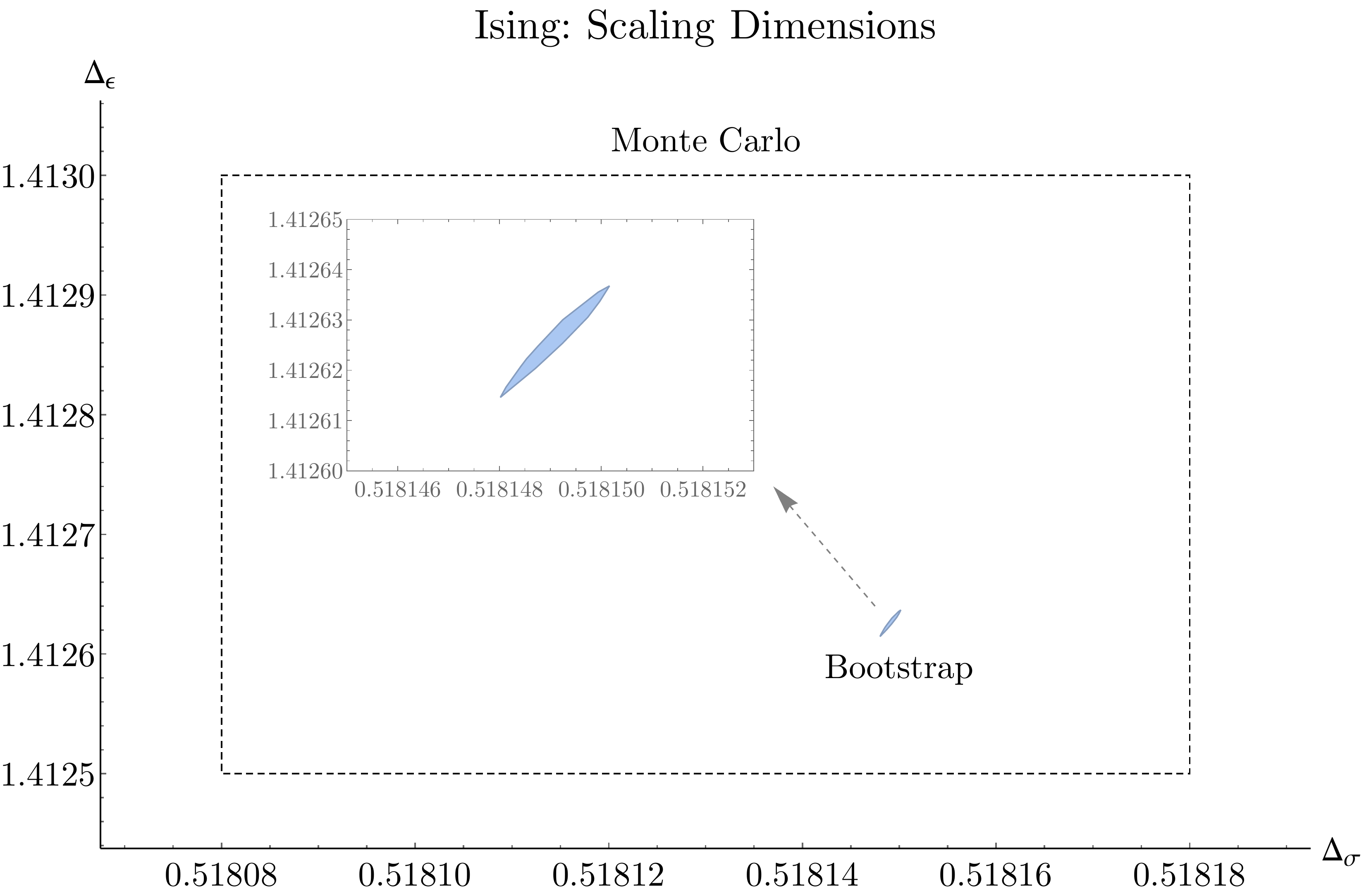}
\caption{\label{fig:IsingIsland} Determination of the leading scaling dimensions in the 3d Ising model from the mixed correlator bootstrap after scanning over the ratio of OPE coefficients $\lambda_{\e\e\e}/\lambda_{\s\s\e}$ and projecting to the $(\De_{\s}, \De_{\e})$ plane (blue region). Here we assume that $\s$ and $\e$ are the only relevant $\mathbb{Z}_2$-odd and $\mathbb{Z}_2$-even scalars, respectively. In this plot we compare to the previous best Monte Carlo determinations~\cite{Hasenbusch:2011yya} (dashed rectangle). This region is computed at $\Lambda=43$.}
\end{figure}

\begin{figure}[t!]
\includegraphics[width=\textwidth]{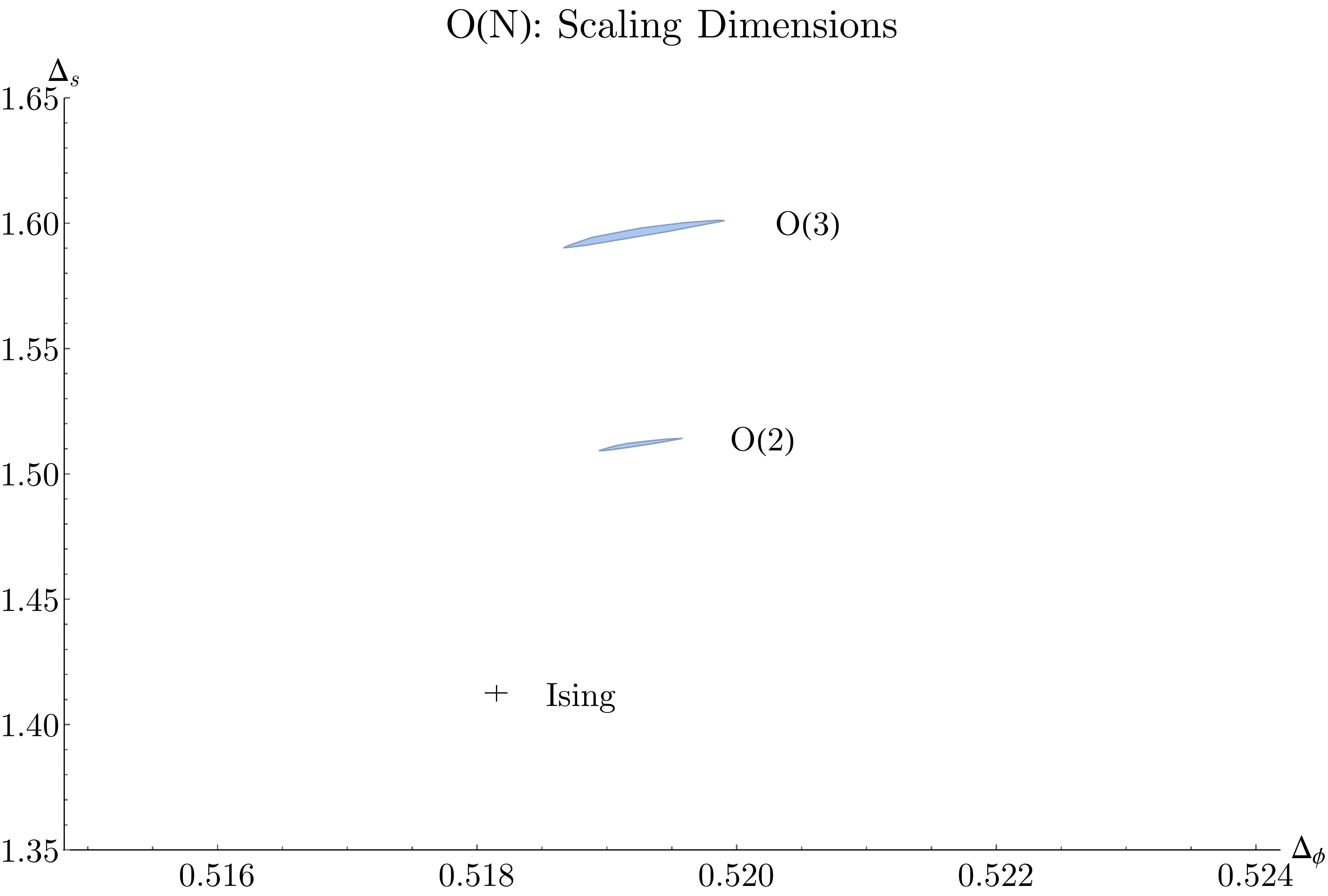}
\caption{\label{fig:ONArchipelago} Allowed islands from the mixed correlator bootstrap for the $O(2)$ and $O(3)$ models after scanning over the ratio of OPE coefficients $\lambda_{sss}/\lambda_{\f\f s}$ and projecting to the ($\De_{\f},\De_{s}$) plane (blue regions). Here we assume that $\phi$ and $s$ are the only relevant scalar operators in their $O(N)$ representations. These islands are computed at $\Lambda=35$. The Ising island is marked with a cross because it is too small to see on the plot.}
\end{figure}
\end{center}

\section{Bootstrap Constraints}
\label{sec:crossing}

\subsection{Ising Model}

We will be studying the conformal bootstrap constraints for 3d CFTs with either a $\Z_2$ or $O(N)$ global symmetry. In the case of a $\Z_2$ symmetry, relevant for the 3d Ising model, we consider all 4-point functions containing the leading $\Z_2$-odd scalar $\sigma$ and leading $\Z_2$-even scalar $\epsilon$. The resulting system of bootstrap equations for $\{\<\s\s\s\s\>, \<\s\s\e\e\>, \<\e\e\e\e\>\}$ was presented in detail in~\cite{Kos:2014bka}. Here we summarize the results. The crossing symmetry conditions for these correlators can be expressed as a set of 5 sum rules:
\be
\label{eq:crossingequationwithv}
0 &=& \sum_{\cO^+} \begin{pmatrix}\l_{\s\s\cO} & \l_{\e\e\cO}\end{pmatrix} \vec{V}_{+,\De,\ell}\begin{pmatrix} \l_{\s\s\cO} \\ \l_{\e\e\cO} \end{pmatrix}+ \sum_{\cO^-} \l_{\s\e\cO}^2 \vec{V}_{-,\De,\ell},
\ee
where $\vec{V}_{-,\De,\ell}$ is a 5-vector and $\vec{V}_{+,\De,\ell}$ is a 5-vector of $2 \times 2$ matrices. The detailed form of $\vec{V}_{\pm}$, describing the contributions of parity even or odd operators $\cO^{\pm}$ in terms of conformal blocks, is given in~\cite{Kos:2014bka}.

In~\cite{Kos:2014bka,Simmons-Duffin:2015qma} we numerically computed the allowed region for $(\De_\s,\De_\e)$ by assuming that $\s$ and $\e$ are the only relevant scalar operators in the spectrum and searching for a functional $\vec{\alpha}$ satisfying the conditions
\be
\label{eq:functionalinequalities}
\begin{pmatrix} 1 & 1\end{pmatrix} \vec \alpha \cdot \vec{V}_{+,0,0} \begin{pmatrix} 1 \\ 1 \end{pmatrix}  &>& 0 ,\quad\textrm{for the identity operator}, \nn\\
\vec \alpha \cdot \vec{V}_{+,\De,\ell} &\succeq &0 ,\quad\textrm{for $\mathbb{Z}_2$-even operators with even spin,} \nn\\
\vec \alpha \cdot \vec{V}_{-,\De,\ell} & \ge &0 ,\quad \textrm{for $\mathbb{Z}_2$-odd operators in the spectrum.} 
\ee
If such a functional can be found, then the assumed values of $(\De_\s,\De_\e)$ are incompatible with unitarity or reflection positivity. In~\cite{Kos:2014bka,Simmons-Duffin:2015qma} we found that this leads to an isolated allowed island in operator dimension space compatible with known values in the 3d Ising model, with a size dependent on the size of the search space for the functional. One can additionally incorporate the constraint $\l_{\s\s\e} = \l_{\s\e\s}$ by only requiring positivity for the combination 
\be\label{eq:OPErelation}
\vec{\alpha} \cdot \left(\vec{V}_{+,\De_{\e},0} + \vec{V}_{-,\De_{\s},0} \otimes \left(\begin{array}{cc}1 &0 \\0 &0\end{array}\right)\right) \succeq 0,
\ee
reducing the size of the island somewhat further.

However, as noted in~\cite{Kos:2014bka}, the condition~(\ref{eq:OPErelation}) is still stronger than necessary. In particular it allows for solutions of crossing containing terms of the form
\be\label{eq:strongversion}
 \sum_{i} \begin{pmatrix}\l_{\s\s i} & \l_{\e\e i}\end{pmatrix}\left(\vec{V}_{+,\De_{\e},0} + \vec{V}_{-,\De_{\s},0} \otimes \left(\begin{array}{cc}1 &0 \\0 &0\end{array}\right)\right) \begin{pmatrix} \l_{\s\s i} \\ \l_{\e\e i} \end{pmatrix},
\ee
where $\begin{pmatrix}\l_{\s\s i} & \l_{\e\e i}\end{pmatrix}$ represent an arbitrary number of (not necessarily aligned) two-component vectors.
If instead we assume that $\s$ and $\e$ are isolated and that there are no other contributions at their scaling dimensions, then we can replace~(\ref{eq:OPErelation}) with the weaker condition
\be\label{eq:OPErelationweaker}
\begin{pmatrix} \cos\theta & \sin\theta \end{pmatrix} \vec{\alpha} \cdot \left(\vec{V}_{+,\De_{\e},0} + \vec{V}_{-,\De_{\s},0} \otimes \left(\begin{array}{cc}1 &0 \\0 &0\end{array}\right)\right) \begin{pmatrix} \cos\theta \\ \sin\theta \end{pmatrix} \geq 0,
\ee
for some unknown angle $\theta \equiv \tan^{-1}(\l_{\e\e\e}/\l_{\s\s\e})$. By scanning over the possible values of $\theta$ and taking the union of the resulting allowed regions (an idea first explored in~\cite{SlavaUnpublished}), we can effectively allow our functional to depend on this unknown ratio and arrive at a smaller allowed region, forbidding solutions to crossing of the uninteresting form~(\ref{eq:strongversion}).

In addition, for any given allowed point in the $(\Delta_\sigma,\Delta_\epsilon,\theta)$ space, we can compute a lower and upper bound on the norm $\lambda_\epsilon \equiv \sqrt{\l_{\s\s\e}^2+\l_{\e\e\e}^2 }$ of the OPE coefficient vector. This is obtained by substituting the conditions \ref{eq:functionalinequalities} with the optimization problem:
\be
\label{eq:functionalinequalitiesOPE}
\text{Maximize} &&\begin{pmatrix} 1 & 1\end{pmatrix} \vec \alpha \cdot \vec{V}_{+,0,0} \begin{pmatrix} 1 &1 \end{pmatrix} \qquad\text{subject to}\nn\\
 \mathcal N& =&\begin{pmatrix} \cos\theta & \sin\theta \end{pmatrix} \vec{\alpha} \cdot \left(\vec{V}_{+,\De_{\e},0} + \vec{V}_{-,\De_{\s},0} \otimes \left(\begin{array}{cc}1 &0 \\0 &0\end{array}\right)\right) \begin{pmatrix} \cos\theta \\ \sin\theta \end{pmatrix},\nn \\ 
\vec \alpha \cdot \vec{V}_{+,\De,\ell} &\succeq &0 ,\quad\textrm{for $\mathbb{Z}_2$-even operators with even spin,} \nn\\
\vec \alpha \cdot \vec{V}_{-,\De,\ell} & \ge&0  ,\quad \textrm{for $\mathbb{Z}_2$-odd operators in the spectrum.} 
\ee
By choosing $\mathcal N=\pm 1$ we can obtain the sought upper and lower bounds:
\be
\mathcal N   \lambda_\epsilon^2 \leq -\begin{pmatrix} 1 & 1\end{pmatrix} \vec \alpha \cdot \vec{V}_{+,0,0} \begin{pmatrix} 1 &1 \end{pmatrix} .
\ee

\subsection{$O(N)$ Models}

Similarly, when there is an $O(N)$ symmetry, we can consider all 4-point functions containing the leading $O(N)$ vector $\phi_i$ and leading $O(N)$ singlet $s$. The resulting system of bootstrap equations for $\{\<\f\f\f\f\>, \<\f\f s s\>, \<ssss\>\}$ was studied in~\cite{Kos:2015mba}, leading to a set of 7 sum rules of the form
\be\label{eq:vectoreq}
0 &=& \sum_{\cO_S,\ell^+} \left(\begin{array}{ccc} \l_{\f\f \cO_S} & \l_{ss \cO_S} \end{array} \right) \vec{V}_{S,\De,\ell} \left( \begin{array}{c} \l_{\f\f \cO_S} \\ \l_{ss \cO_S} \end{array} \right) +\sum_{\cO_T,\ell^+} \l_{\f\f \cO_T}^2 \vec{V}_{T,\De,\ell}\nn\\
&&+ \sum_{\cO_A,\ell^-} \l_{\f\f \cO_A}^2 \vec{V}_{A,\De,\ell}+ \sum_{\cO_V,\ell^\pm} \l_{\f s \cO_V}^2 \vec{V}_{V,\De,\ell},
\ee
where $\vec{V}_{T}, \vec{V}_A, \vec{V}_V$ are 7-dimensional vectors corresponding to different choices of correlators and tensor structures and $\vec{V}_{S}$ is a 7-vector of $2 \times 2$ matrices. The functions $\vec{V}_S, \vec{V}_{T}, \vec{V}_A, \vec{V}_V$ describe the contributions from singlets $\cO_S$, symmetric tensors $\cO_T$, anti-symmetric tensors $\cO_A$, and vectors $\cO_V$, and are defined in detail in~\cite{Kos:2015mba}.

To rule out an assumption on the spectrum, we will look for a functional satisfying the generic conditions
\be\label{eq:functionalconditionsON}
& \left(\begin{array}{ccc} 1 & 1 \end{array}\right) \vec\alpha\cdot \vec{V}_{S,0,0} \left( \begin{array}{c} 1 \\ 1 \end{array} \right) \geq 0,&\text{for the identity operator},\nn\\
&\vec\alpha\cdot \vec{V}_{T,\De,\ell} \geq 0,&\text{for traceless symetric tensors with $\ell$ even},\label{eq:functional-symtensor} \nn\\
&\vec\alpha\cdot \vec{V}_{A,\De,\ell}  \geq 0,&\text{for antisymmetric tensors with $\ell$ odd},\nn\\
&\vec\alpha\cdot \vec{V}_{V,\De,\ell} \geq 0,&\text{for $O(N)$ vectors with any $\ell$}\nn\label{eq:functional-fundamental},\\
&\vec\alpha\cdot \vec{V}_{S,\De,\ell} \succeq 0,&\text{for singlets with $\ell$ even},\label{eq:functional-singlet}
\ee
where we take these constraints to hold for scalar singlets and vectors with $\De \geq 3$, symmetric tensors with $\De \geq 1$, and all operators with spin satisfying the unitarity bound $\De \geq \ell + 1$. Similar to the previous section, we will additionally allow for the contributions of the isolated operators $\phi_i$ and $s$ by imposing the condition
\be
\begin{pmatrix} \cos\theta_N & \sin\theta_N \end{pmatrix} \vec\alpha\cdot \left(\vec{V}_{S,\Delta_s,0} + \vec{V}_{V,\Delta_\phi,0} \otimes \left(\begin{array}{cc}1 &0 \\0 &0\end{array}\right)\right) \begin{pmatrix} \cos\theta_N \\ \sin\theta_N \end{pmatrix}  \geq 0
\ee
and scanning over the unknown angle $\theta_N \equiv \tan^{-1}(\l_{sss} / \l_{ss\f})$.

Similarly to the previous section, for any allowed point in $(\Delta_\phi,\Delta_\s,\theta)$ space, we can compute a lower and upper bound on the norm $\lambda_s \equiv \sqrt{\l_{\phi\phi s}^2+\l_{sss}^2 }$. This is obtained by substituting the conditions \ref{eq:functionalconditionsON} with:
\be
\label{eq:functionalinequalitiesONOPE}
\text{Maximize} &&\begin{pmatrix} 1 & 1\end{pmatrix} \vec \alpha \cdot \vec{V}_{S,0,0} \begin{pmatrix} 1 &1 \end{pmatrix} \qquad \text{subject to}\nn\\
 \mathcal N& =&\begin{pmatrix} \cos\theta_N & \sin\theta_N \end{pmatrix} \vec{\alpha} \cdot \left(\vec{V}_{S,\De_{s},0} + \vec{V}_{-,\De_{s},0} \otimes \left(\begin{array}{cc}1 &0 \\0 &0\end{array}\right)\right) \begin{pmatrix} \cos\theta_N \\ \sin\theta_N \end{pmatrix},\nn \\ 
\vec\alpha\cdot \vec{V}_{T,\De,\ell} &\geq 0,&\text{for traceless symetric tensors with $\ell$ even},\label{eq:functional-symtensor} \nn\\
\vec\alpha\cdot \vec{V}_{A,\De,\ell}  &\geq 0,&\text{for antisymmetric tensors with $\ell$ odd},\nn\\
\vec\alpha\cdot \vec{V}_{V,\De,\ell} &\geq 0,&\text{for $O(N)$ vectors with any $\ell$}\nn\label{eq:functional-fundamental},\\
\vec\alpha\cdot \vec{V}_{S,\De,\ell} &\succeq 0, &\text{for singlets with $\ell$ even}.\label{eq:functional-singlet} 
\ee

\section{Results}
\label{sec:results}

\begin{figure}[t!]
\includegraphics[width=\textwidth]{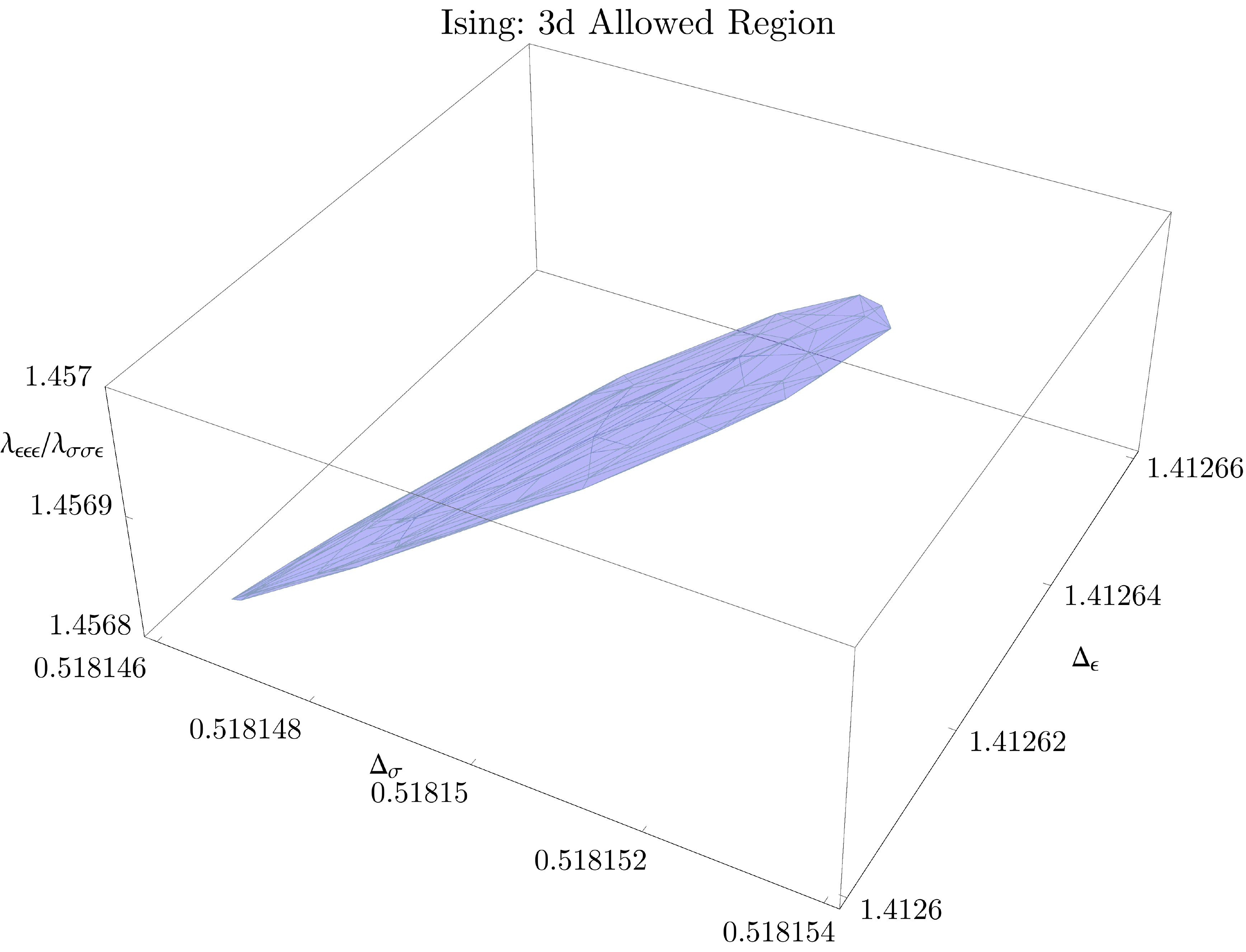}
\caption{\label{fig:3dIsingIsland} Determination of the leading scaling dimensions $(\De_{\s},\De_{\e})$ and the OPE coefficient ratio $\l_{\e\e\e}/\l_{\s\s\e}$ in the 3d Ising model from the mixed correlator bootstrap (blue region). This region is computed at $\Lambda=43$. }
\end{figure}
\begin{figure}[t!]
\includegraphics[width=\textwidth]{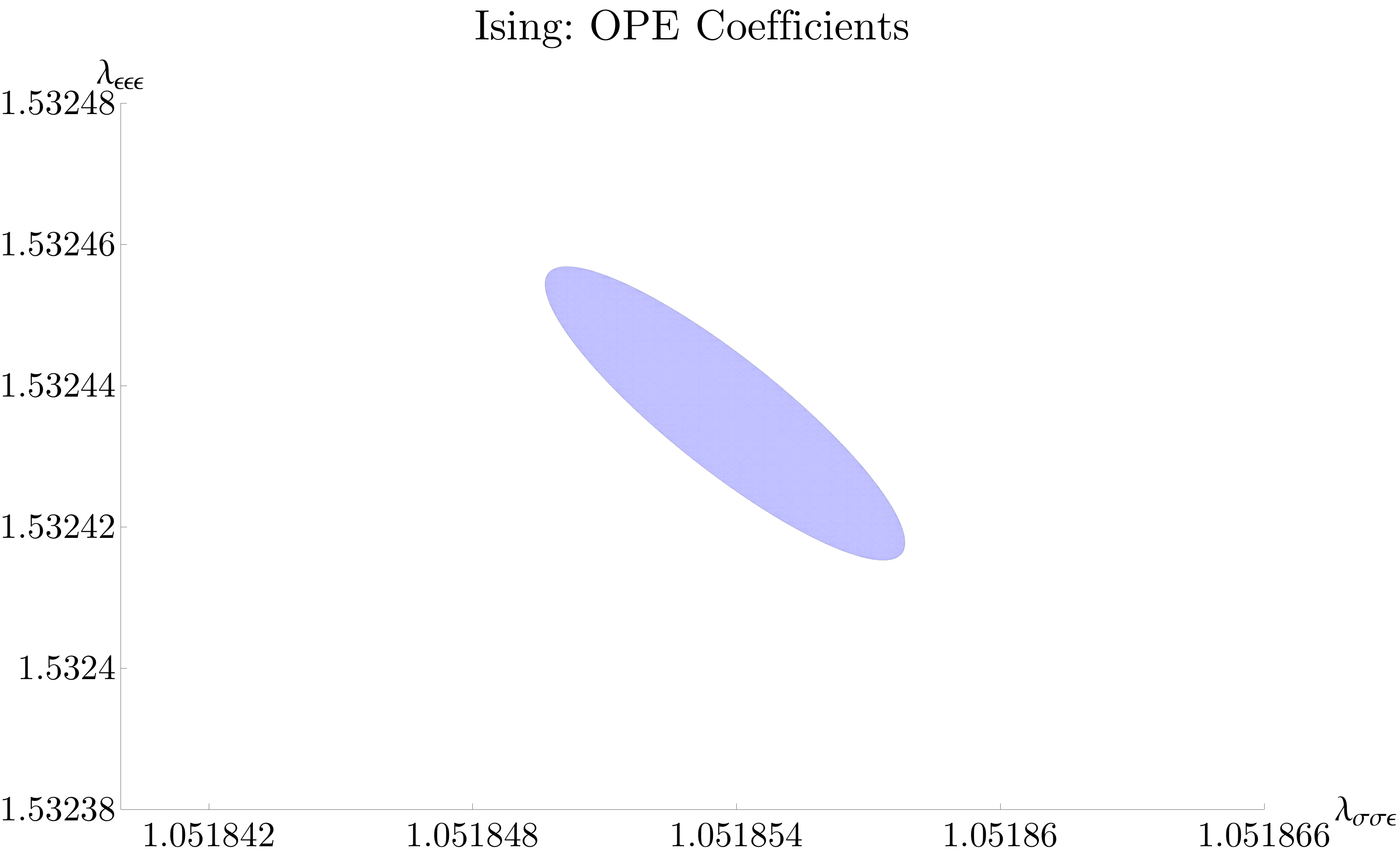}
\caption{\label{fig:OPEIsland} Determination of the leading OPE coefficients in the 3d Ising model from the conformal bootstrap (blue region). This region was obtained by computing upper and lower bounds on the OPE coefficient magnitude at $\Lambda=27$, for points in the allowed region of figure~\ref{fig:3dIsingIsland}.}
\end{figure}
\begin{figure}[t!]
\includegraphics[width=\textwidth]{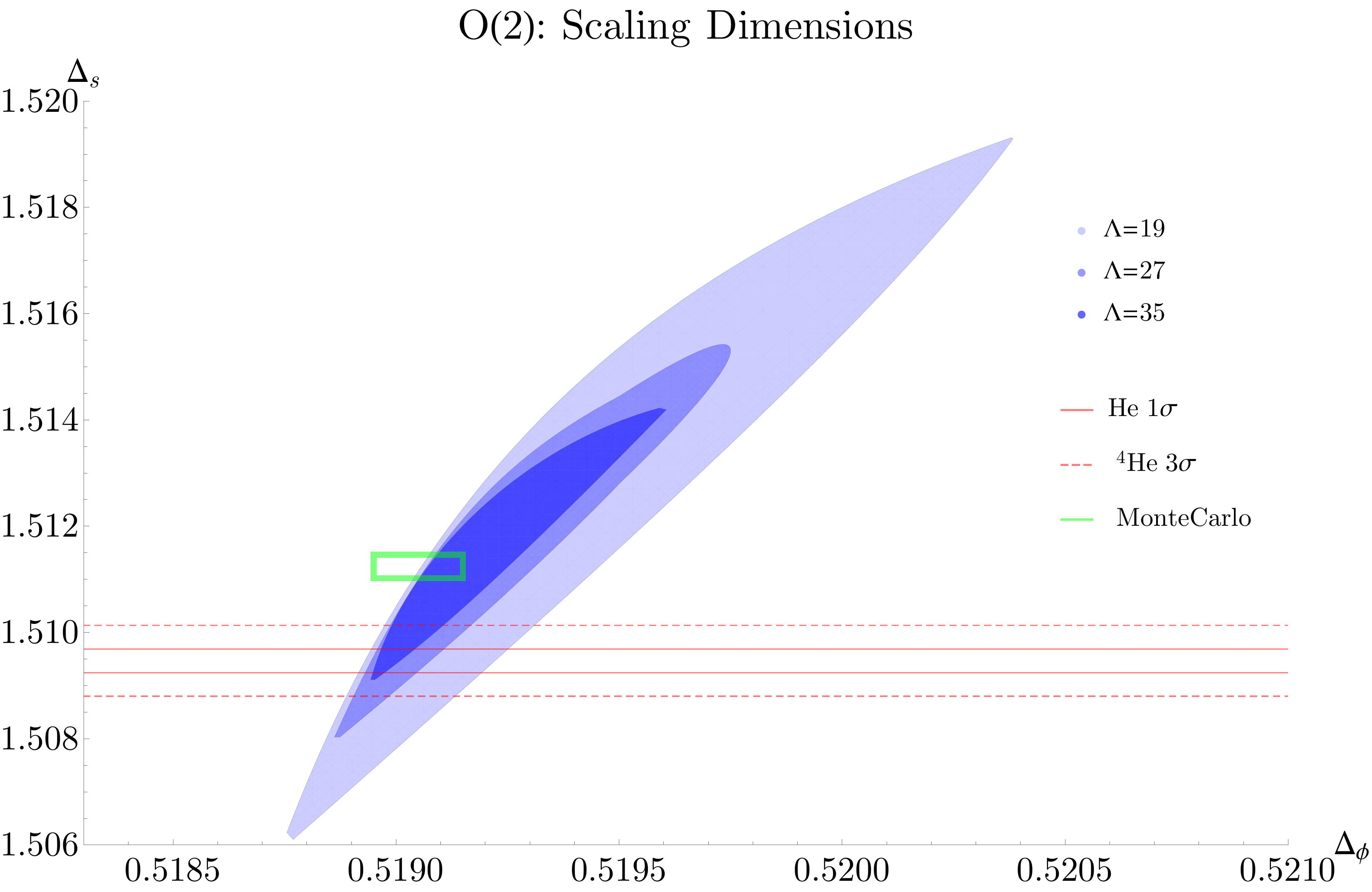}
\caption{\label{fig:O2Island} Allowed islands from the mixed correlator bootstrap for $N=2$ after scanning over the OPE coefficient ratio $\lambda_{sss}/\lambda_{\f\f s}$ and projecting to the $(\De_{\f}, \De_s)$ plane (blue regions). Here we assumed that $\phi$ and $s$ are the only relevant operators in their $O(N)$ representations. These islands are computed at $\Lambda=19,27,35$. The green rectangle shows the Monte Carlo determination from~\cite{Campostrini:2006ms}, while the horizontal lines show the $1\sigma$ (solid) and $3\sigma$ (dashed) confidence intervals from experiment~\cite{Lipa:2003zz}.}
\end{figure}
\begin{figure}[t!]
\includegraphics[width=\textwidth]{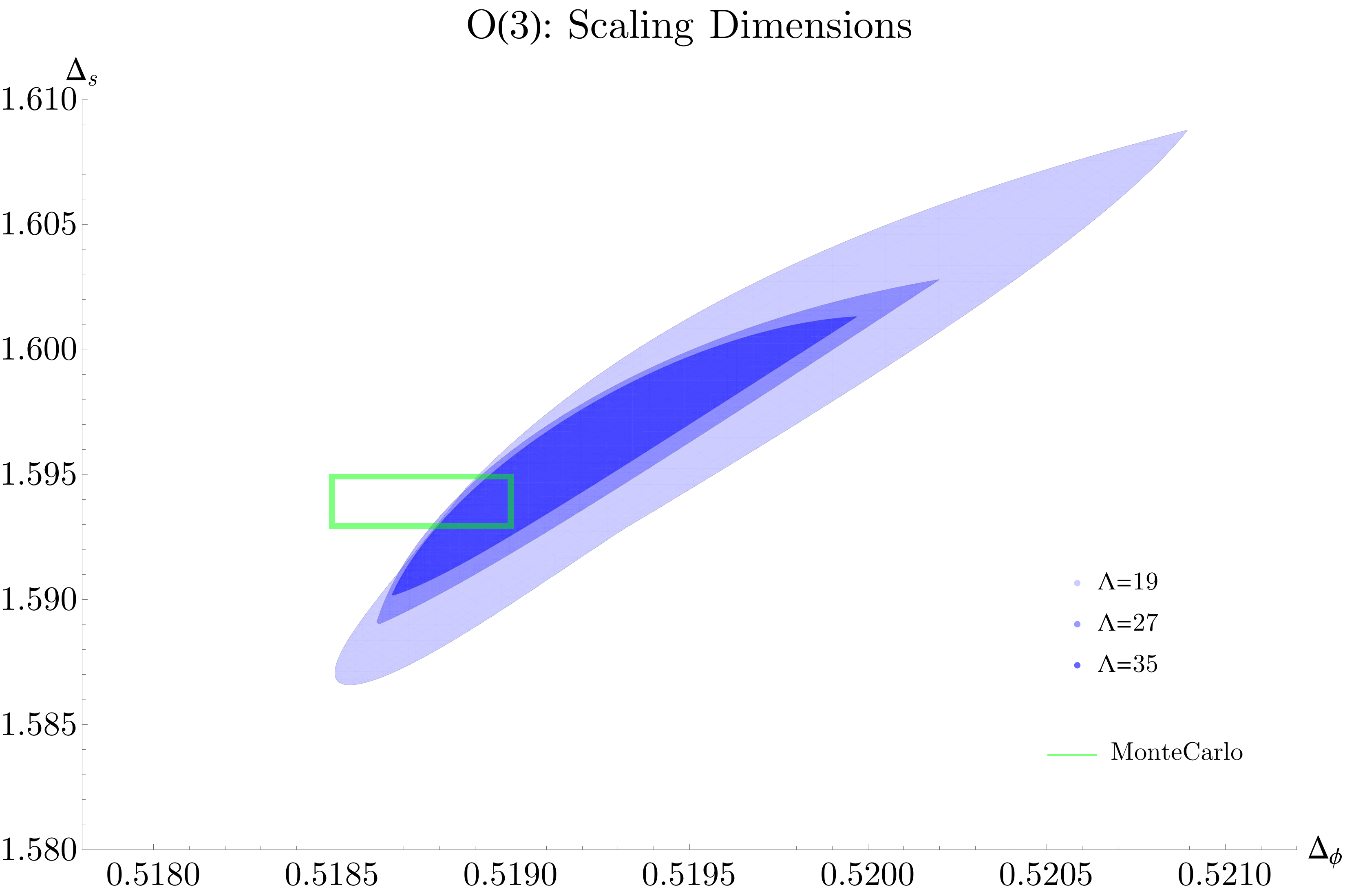}
\caption{\label{fig:O3Island} Allowed islands from the mixed correlator bootstrap for $N=3$ after scanning over the OPE coefficient ratio $\lambda_{sss}/\lambda_{\f\f s}$ and projecting to the $(\De_{\f}, \De_s)$ plane (blue regions). Here we assumed that $\phi$ and $s$ are the only relevant operators in their $O(N)$ representations.  These islands are computed at $\Lambda=19,27,35$. The green rectangle shows the Monte Carlo determination from~\cite{Campostrini:2002ky}.}
\end{figure}

As shown in figures~\ref{fig:IsingIsland} and~\ref{fig:3dIsingIsland},\footnote{In the plots in this work we show smooth curves that have been fit to the computed points. The precise shape of the boundary is subject to an error which is at least an order of magnitude smaller than the quoted error bars.} we have used this procedure to determine the scaling dimensions and OPE coefficient ratio in the 3d Ising model to high precision at $\Lambda=43$,\footnote{The functional $\vec\alpha$ we search for is given as a linear combination of derivatives. The parameter $\Lambda$ limits the highest order derivative that can appear in the functional $\vec\alpha$. See~\cite{Kos:2015mba} for the exact definition of the parameter $\Lambda$.} giving
\be
\label{eq:Isingdeterminations}
\Delta_\s &=& 0.5181489(10),\\
\Delta_\e &=& 1.412625(10),\\
\lambda_{\e\e\e}/\lambda_{\s\s\e} &=& 1.456889(50).
\ee
We have also computed bounds on the magnitude of the leading OPE coefficients $\lambda_{\e}$ at $\Lambda=27$ over this allowed region, with the result shown in figure~\ref{fig:OPEIsland}. These determinations yield the values
\be
\label{eq:Isingopes}
\lambda_{\s\s\e} &=& 1.0518537(41),\\
\lambda_{\e\e\e} &=& 1.532435(19).
\ee
Our determination of $\lambda_{\e\e\e}$ is consistent with the estimate $1.45\pm 0.3$ obtained via Monte Carlo methods in \cite{Caselle:2015csa}.\footnote{We disagree slightly with the determination in \cite{Costagliola:2015ier}.}  An application of $\lambda_{\e\e\e}$ is in calculating the properties of the 3d Ising model in the presence of quenched disorder in the interaction strength of neighboring spins \cite{DisorderedIsing}.

In figure~\ref{fig:ONArchipelago} we show similar islands for the leading vector and singlet operators in the $O(2)$ and $O(3)$ models, all computed at $\Lambda = 35$. We show the zoom in of these regions as well as the regions at $\Lambda=19,27$ in figures~\ref{fig:O2Island} and~\ref{fig:O3Island}.  Once the angle $\theta_N$ has been computed at $\Lambda=35$, we determine the OPE coefficients $(\lambda_{\f\f s}, \lambda_{sss})$ by bounding the magnitude $\lambda_{s}$ at $\Lambda=27$.  The final error in the OPE coefficients comes mostly from the angle, which is why we use a lower value of $\Lambda$ for the magnitude.

For the $O(2)$ model, the resulting dimensions and OPE coefficients are
\be
\label{eq:O2determinations}
\Delta_{\f} &=& 0.51926(32),\\
\Delta_{s} &=& 1.5117(25),\\
\lambda_{sss}/\lambda_{\f\f s} &=& 1.205(9),\\
\lambda_{\f\f s} &=& 0.68726(65),\\
\lambda_{sss} &=& 0.8286(60).
\ee
%Computing upper and lower bounds on the magnitude of the OPE coefficients at $\Lambda=27$ then gives
%\be
%\ee

A similar computation for the $O(3)$ model gives
\be
\label{eq:O3determinations}
\Delta_{\f} &=& 0.51928(62),\\
\Delta_{s} &=& 1.5957(55),\\
\lambda_{sss}/\lambda_{\f\f s} &=& 0.953(25),\\
\lambda_{\f\f s} &=& 0.5244(11),\\
\lambda_{sss} &=& 0.499(12).
\ee

In the $O(2)$ plot we compare to both the Monte Carlo determinations of~\cite{Campostrini:2006ms} and the re-analysis of the experimental ${}^4$He data of~\cite{Lipa:2003zz}, currently in $\sim 8\s$ tension. Our result is easily compatible with~\cite{Campostrini:2006ms} while it has started to exclude the lower part of the $3\s$ allowed region reported in~\cite{Lipa:2003zz}. Based on a na\"ive extrapolation to a higher derivative cutoff $\Lambda$, it seems plausible that the bootstrap result will eventually fully exclude the reported result of~\cite{Lipa:2003zz}. If this occurs, we would attribute the discrepancy to the fact that the fit performed in~\cite{Lipa:2003zz} has a sizable sensitivity to which subleading contributions to the heat capacity are included, as can be seen in table II of~\cite{Lipa:2003zz}. It is therefore plausible to us that the experimental uncertainty in the extraction of the critical exponent $\alpha$ should be larger than the reported error bars.

\section{Conclusions}
\label{sec:conclusions}

In this work we imposed the uniqueness of the relevant singlet operator appearing in the conformal block decomposition of $\<\f\f\f\f\>$, $\<\f\f s s\>$, and $\<ssss\>$ in the Ising and $O(N)$ models.\footnote{To unify the discussion we use the $O(N)$ notation to denote operator dimension and OPE coefficients, with the obvious dictionary to translate to the Ising model: $\phi \rightarrow \sigma$, $s \rightarrow \epsilon$.}  The absence of degeneracies is a natural restriction to impose on the CFT spectrum. It requires a modified numerical approach because the standard mixed correlator analysis used in previous works \cite{Kos:2014bka,Simmons-Duffin:2015qma,Kos:2015mba,Lemos:2015awa,Behan:2016dtz,Nakayama:2016jhq} secretly allows for more general solutions of crossing symmetry that violate this assumption. 

We implement this new constraint by scanning over the ratio of OPE coefficients $\l_{sss}/\l_{\f\f s}$. By forbidding uninteresting solutions of crossing we further restricted the allowed region in the  $(\De_\phi,\De_s)$ plane. This results in a new precise determination of Ising critical exponents, almost two orders of magnitude better than the best Monte Carlo estimate~\cite{Hasenbusch:2011yya}. We also improved on our previous determinations for $O(2)$ and $O(3)$, although Monte Carlo results for operator dimensions remain more precise in those cases. (The bootstrap allows much more precise determinations of OPE coefficients.) Nevertheless, for $O(2)$, we saw indications that the conformal bootstrap disfavors the commonly-quoted value extracted from experimental ${}^4$He data in the analysis of~\cite{Lipa:2003zz}.

For the sake of completeness we also report qualitative results of attempts to reduce the size of the allowed regions by imposing additional assumptions. One natural ingredient not exploited so far is the constraint that the energy momentum tensor appears with the same central charge in all correlators. Enforcing this also requires imposing a gap between $\Delta_T = 3$ and the dimension of the next spin two operator, $\Delta_{T'}=3+\delta$. The net effect is a non-negligible shrinking of the size of the $O(2)$ island, but unfortunately it only carves out the upper right region of the island, leaving the rest essentially untouched. The effect is also independent of the value of the gap as long as $0.2\le\delta \le 1$. Finally, we found that the lower left endpoint of the $O(2)$ island is controlled by the gap between $\Delta_s$ and the dimension of the next singlet scalar $\Delta_{s'}$; however only when we assume $\Delta_{s'}>3.7$ do we start changing the size of the $O(2)$ island. This is not surprising since the expected value from MonteCarlo is $\Delta_{s'} = 3.785(20)$ \cite{Campostrini:2006ms}. In order to keep the discussion general we decided not to push further in this direction.

As a byproduct of our analysis, we also obtained precise determinations of the OPE coefficients $(\l_{\phi\phi s},\l_{sss})$. While the latter is here computed for the first time, the former was already estimated for the Ising model  in \cite{El-Showk:2014dwa}, using again a bootstrap approach. There the value $\left|\l_{\phi\phi s}\right|=1.05183(86)$ was extracted, which should be compared with the result in (\ref{eq:Isingopes}). The two determinations are fully compatible, despite the methods used to obtain the two estimates being somewhat different, both in the theoretical and numerical approach to the conformal bootstrap. The present work uses mixed correlators, it translates the crossing constraints into a semidefinite programming problem, and rules out unfeasible points in the CFT parameter space. The work of \cite{El-Showk:2014dwa}, instead, used a linear programming algorithm to solve the crossing equations directly under the assumption that the 3d Ising model is the 3d CFT which locally minimizes the central charge. The agreement of these methods is a further triumph of the numerical bootstrap.

The new ingredient studied in this work represents a further step in the numerical development of the conformal bootstrap. It not only further reduces the size of the allowed parameter space, but it also provides rigorous information on OPE coefficients. It will be interesting to investigate the effect of scanning over relative OPE coefficients in other situations where the bootstrap seems to be successful, both in known theories such as $\cN=4$ supersymmetric Yang-Mills theory~\cite{Beem:2013qxa}, the 6d (2,0) SCFTs~\cite{Beem:2015aoa}, and the conformal window of QCD~\cite{Iha:2016ppj}, as well as in studies of the mysterious features that have appeared in the 4d $\cN=1$~\cite{Poland:2011ey,Poland:2015mta} and 3d fermion~\cite{Iliesiu:2015qra} bootstrap that may signal the existence of new islands in the ocean of CFTs.

\section*{Acknowledgements}
We thank Zohar Komargodski and Slava Rychkov for discussions.
The work of DSD is supported by DOE grant number DE-SC0009988 and a William D. Loughlin Membership at the Institute for Advanced Study. The work of DP and FK is supported by NSF grant 1350180. DP is additionally supported by a Martin A. and Helen Chooljian Founders' Circle Membership at the Institute for Advanced Study.  The computations in this paper were run on the Bulldog computing clusters supported by the facilities and staff of the Yale University Faculty of Arts and Sciences High Performance Computing Center, on the Hyperion computing cluster supported by the School of Natural Sciences Computing Staff at the Institute for Advanced Study, and on the CERN cluster.

\appendix

\section{Implementation Details}
\label{sec:appA}

Using the techniques described in the main text, we can set up a semidefinite program to determine whether a triple $(\De_\s,\De_\e,\theta)$ is allowed. (In this discussion, we focus on the Ising model for simplicity.)  Our choices and parameters for solving the semidefinite program are identical to those quoted in \cite{Kos:2015mba}.   To actually determine $(\De_\s,\De_\e,\theta)$ in the Ising model, we must make a 3d exclusion plot at successively larger values of $\Lambda$.  We proceed as follows:

\begin{itemize}

\item We first choose a relatively small value $\Lambda=\Lambda_0$.  (For us, $\Lambda_0=11$.)  Since we roughly know the 2d projection of the 3d Island from previous work \cite{Simmons-Duffin:2015qma}, we begin by choosing some points $(\De_\s,\De_\e)$ in the 2d island and performing a 1d scan over $\theta$.  If we're lucky, this gives at least one point $p_0$ in the 3d island.

\item By scanning over a 3d grid near $p_0$, we determine the rough shape $S_{\Lambda_0}$ of the 3d island. 

\item The island shrinks in an approximately self-similar way as $\Lambda$ is increased.  Once we know the shape $S_{\Lambda_0}$, we find an affine transformation $T_{\Lambda_0}:S_{\Lambda_0}\to [-1,1]^3$ such that $S_{\Lambda=19}$ becomes approximately spherical, with large volume in $[-1,1]^3$.  $T_{\Lambda_0}$ gives a useful set of coordinates for a neighborhood of $S_{\Lambda_0}$.  These coordinates are much better than $(\De_\s,\De_\e,\theta)$, because $S_{\Lambda_0}$ is extremely elongated and flat in $(\De_\s,\De_\e,\theta)$ space (figure~\ref{fig:3dIsingIsland}).  It is helpful to choose $T_{\Lambda_0}$ so that the plane $\De_\e-\De_\s=0$ is parallel to two of the axes in $[-1,1]^3$.  This ensures that a grid-based scan over $[-1,1]^3$ involves only a small number of values of $\De_\e-\De_\s$, which means we must compute fewer tables of conformal blocks.  This is the 3d generalization of the trick mentioned in \cite{Kos:2014bka}.

\item Now that we have a better reference frame for $S_{\Lambda_0}$, our job is easier.  We increase $\Lambda_0\to \Lambda_1$ and determine a point $p_{1}\in S_{\Lambda_1}$ using a rough scan. We then determine the boundary of $S_{\Lambda_1}$ by performing a binary search in the radial direction away from $p_1$, in the $T_{\Lambda_0}$ coordinates.  For the angular directions, we choose the vertices and edge-midpoints of an icosahedron centered at $p_1$, oriented so that $\De_\e-\De_\s$ takes as few values as possible during the search.  To get a higher resolution picture of $S_{\Lambda_1}$, we can pick a few more points in the interior and perform radial binary searches away from those points as well.  Once we know $S_{\Lambda_1}$ we choose a new $T_{\Lambda_1}:S_{\Lambda_1}\to [-1,1]^3$.

\item We now iterate the previous step to increase $\Lambda_1\to \Lambda_2\to \Lambda_3\dots$.  After a few iterations, we can predict the point the islands are shrinking towards, removing the need for a scan at each stage.  We take $\Lambda_0=11,\Lambda_1=19,\Lambda_2=27,\Lambda_3=35,\Lambda_4=43$.

\end{itemize}

\begin{figure}
\begin{center}
\includegraphics[width=0.8\textwidth]{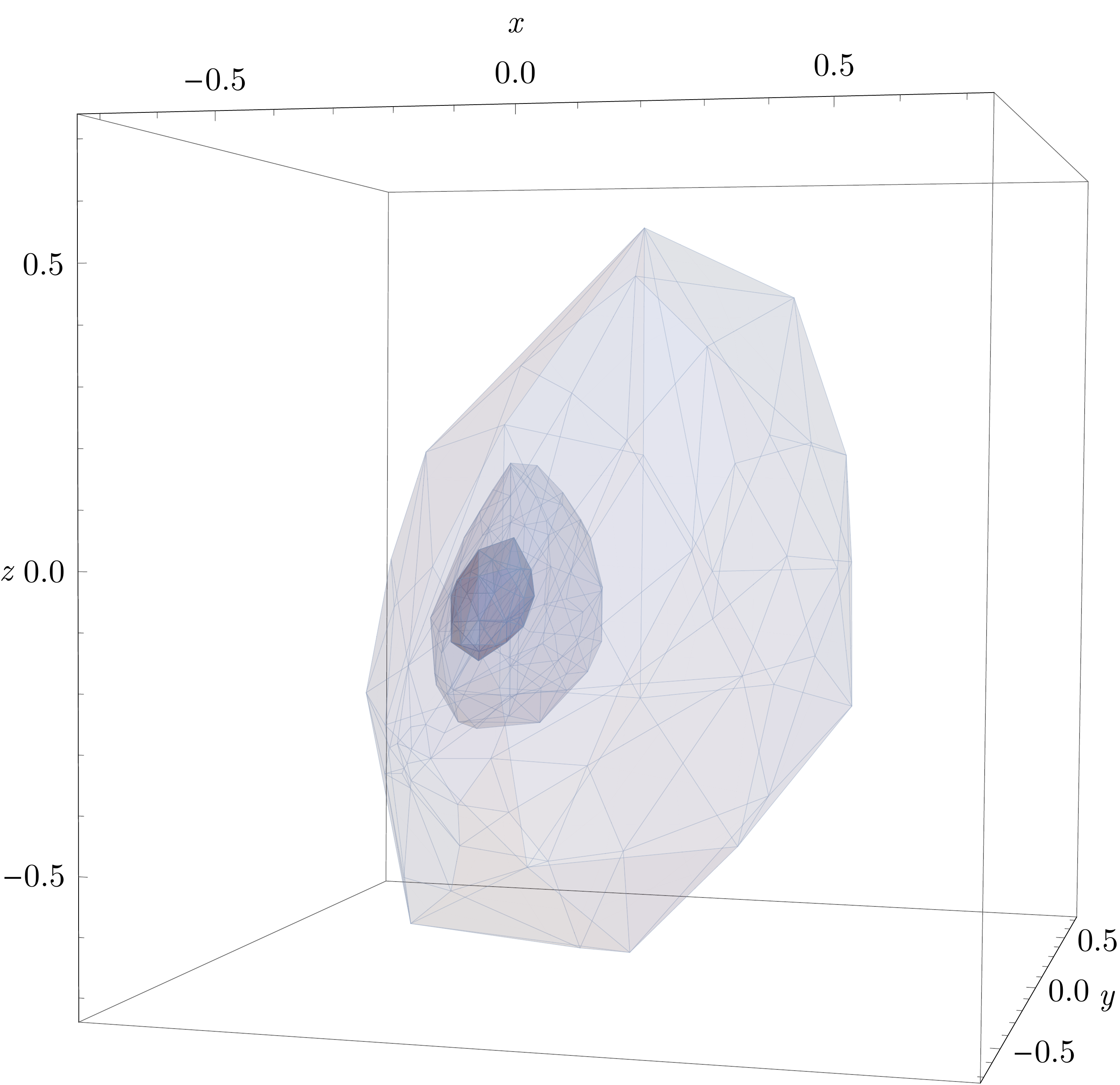}
\end{center}
\caption{\label{fig:affineislands} Images of the 3d islands $\{S_{\Lambda=27}, S_{\Lambda=35}, S_{\Lambda=43}\}$ under the map $T_{\Lambda=27}$, where $T_{\Lambda=27}^{-1}$ is given in~(\ref{eq:ttwentyseveninverse}).}
\end{figure}

As an example, in the 3d Ising model, the inverse map $T_{\Lambda=27}^{-1}$ is given by
\be
\label{eq:ttwentyseveninverse}
\begin{pmatrix}\De_\s \\ \De_\e \\ \th\end{pmatrix}
&=& T_{\Lambda=27}^{-1}\begin{pmatrix}
x\\
y\\
z
\end{pmatrix}\nn\\
&=&
\begin{pmatrix}
2.76988363\cdot10^{-6} & -6.95457153\cdot 10^{-7} & -9.83371791\cdot 10^{-6} \\
2.76988363\cdot10^{-6} & -6.95457153\cdot 10^{-7} & -9.39012428\cdot 10^{-5}  \\
-2.66723434\cdot10^{-5} & -2.70007022\cdot 10^{-6} & -5.48817612\cdot 10^{-5}
\end{pmatrix}\begin{pmatrix} x \\ y \\ z \end{pmatrix}
\nn\\
&& + \begin{pmatrix}
0.51814922\\
1.41261837\\
0.96924816
\end{pmatrix},
\ee
where $(x,y,z)\in [-1,1]^3$.  Note that $\De_\e-\De_\s$ is a function of $z$ alone, which is helpful for reducing the number of tables of conformal blocks needed for scans. The images of the 3d islands $\{S_{\Lambda=27}, S_{\Lambda=35}, S_{\Lambda=43}\}$ under $T_{\Lambda=27}$ are shown in figure~\ref{fig:affineislands}.

\clearpage
\bibliography{Biblio}{}

\providecommand{\href}[2]{#2}\begingroup\raggedright\begin{thebibliography}{10}

\bibitem{Ferrara:1973yt}
S.~Ferrara, A.~F. Grillo, and R.~Gatto, ``{Tensor representations of conformal
  algebra and conformally covariant operator product expansion},''
\href{http://dx.doi.org/10.1016/0003-4916(73)90446-6}{{\em Annals Phys.}
  {\bfseries 76} (1973) 161--188}.
%%CITATION = APNYA,76,161;%%.

\bibitem{Polyakov:1974gs}
A.~M. Polyakov, ``{Nonhamiltonian approach to conformal quantum field
  theory},''
{\em Zh. Eksp. Teor. Fiz.} {\bfseries 66} (1974) 23--42.
%%CITATION = ZETFA,66,23;%%.

\bibitem{ElShowk:2012ht}
S.~El-Showk, M.~F. Paulos, D.~Poland, S.~Rychkov, D.~Simmons-Duffin, and
  A.~Vichi, ``{Solving the 3D Ising Model with the Conformal Bootstrap},''
  \href{http://dx.doi.org/10.1103/PhysRevD.86.025022}{{\em Phys.Rev.}
  {\bfseries D86} (2012) 025022},
\href{http://arxiv.org/abs/1203.6064}{{\ttfamily arXiv:1203.6064 [hep-th]}}.
%%CITATION = ARXIV:1203.6064;%%.

\bibitem{El-Showk:2014dwa}
S.~El-Showk, M.~F. Paulos, D.~Poland, S.~Rychkov, D.~Simmons-Duffin, {\em
  et~al.}, ``{Solving the 3d Ising Model with the Conformal Bootstrap II.
  c-Minimization and Precise Critical Exponents},''
  \href{http://dx.doi.org/10.1007/s10955-014-1042-7}{{\em J.Stat.Phys.}
  {\bfseries 157} (2014) 869},
\href{http://arxiv.org/abs/1403.4545}{{\ttfamily arXiv:1403.4545 [hep-th]}}.
%%CITATION = ARXIV:1403.4545;%%.

\bibitem{Kos:2014bka}
F.~Kos, D.~Poland, and D.~Simmons-Duffin, ``{Bootstrapping Mixed Correlators in
  the 3D Ising Model},'' \href{http://dx.doi.org/10.1007/JHEP11(2014)109}{{\em
  JHEP} {\bfseries 1411} (2014) 109},
\href{http://arxiv.org/abs/1406.4858}{{\ttfamily arXiv:1406.4858 [hep-th]}}.
%%CITATION = ARXIV:1406.4858;%%.

\bibitem{Simmons-Duffin:2015qma}
D.~Simmons-Duffin, ``{A Semidefinite Program Solver for the Conformal
  Bootstrap},''
\href{http://arxiv.org/abs/1502.02033}{{\ttfamily arXiv:1502.02033 [hep-th]}}.
%%CITATION = ARXIV:1502.02033;%%.

\bibitem{Gliozzi:2013ysa}
F.~Gliozzi, ``{More constraining conformal bootstrap},''
  \href{http://dx.doi.org/10.1103/PhysRevLett.111.161602}{{\em Phys.Rev.Lett.}
  {\bfseries 111} (2013) 161602},
\href{http://arxiv.org/abs/1307.3111}{{\ttfamily arXiv:1307.3111}}.
%%CITATION = ARXIV:1307.3111;%%.

\bibitem{Gliozzi:2014jsa}
F.~Gliozzi and A.~Rago, ``{Critical exponents of the 3d Ising and related
  models from Conformal Bootstrap},''
  \href{http://dx.doi.org/10.1007/JHEP10(2014)042}{{\em JHEP} {\bfseries 1410}
  (2014) 42},
\href{http://arxiv.org/abs/1403.6003}{{\ttfamily arXiv:1403.6003 [hep-th]}}.
%%CITATION = ARXIV:1403.6003;%%.

\bibitem{Kos:2015mba}
F.~Kos, D.~Poland, D.~Simmons-Duffin, and A.~Vichi, ``{Bootstrapping the O(N)
  Archipelago},'' \href{http://dx.doi.org/10.1007/JHEP11(2015)106}{{\em JHEP}
  {\bfseries 11} (2015) 106},
\href{http://arxiv.org/abs/1504.07997}{{\ttfamily arXiv:1504.07997 [hep-th]}}.
%%CITATION = ARXIV:1504.07997;%%.

\bibitem{Kos:2013tga}
F.~Kos, D.~Poland, and D.~Simmons-Duffin, ``{Bootstrapping the $O(N)$ vector
  models},'' \href{http://dx.doi.org/10.1007/JHEP06(2014)091}{{\em JHEP}
  {\bfseries 1406} (2014) 091},
\href{http://arxiv.org/abs/1307.6856}{{\ttfamily arXiv:1307.6856 [hep-th]}}.
%%CITATION = ARXIV:1307.6856;%%.

\bibitem{Nakayama:2014yia}
Y.~Nakayama and T.~Ohtsuki, ``{Five dimensional $O(N)$-symmetric CFTs from
  conformal bootstrap},''
  \href{http://dx.doi.org/10.1016/j.physletb.2014.05.058}{{\em Phys.Lett.}
  {\bfseries B734} (2014) 193--197},
\href{http://arxiv.org/abs/1404.5201}{{\ttfamily arXiv:1404.5201 [hep-th]}}.
%%CITATION = ARXIV:1404.5201;%%.

\bibitem{Lemos:2015awa}
M.~Lemos and P.~Liendo, ``{Bootstrapping $ \mathcal{N}=2 $ chiral
  correlators},'' \href{http://dx.doi.org/10.1007/JHEP01(2016)025}{{\em JHEP}
  {\bfseries 01} (2016) 025},
\href{http://arxiv.org/abs/1510.03866}{{\ttfamily arXiv:1510.03866 [hep-th]}}.
%%CITATION = ARXIV:1510.03866;%%.

\bibitem{Behan:2016dtz}
C.~Behan, ``{PyCFTBoot: A flexible interface for the conformal bootstrap},''
\href{http://arxiv.org/abs/1602.02810}{{\ttfamily arXiv:1602.02810 [hep-th]}}.
%%CITATION = ARXIV:1602.02810;%%.

\bibitem{Nakayama:2016jhq}
Y.~Nakayama and T.~Ohtsuki, ``{Conformal Bootstrap Dashing Hopes of Emergent
  Symmetry},''
\href{http://arxiv.org/abs/1602.07295}{{\ttfamily arXiv:1602.07295
  [cond-mat.str-el]}}.
%%CITATION = ARXIV:1602.07295;%%.

\bibitem{Iha:2016ppj}
H.~Iha, H.~Makino, and H.~Suzuki, ``{Upper bound on the mass anomalous
  dimension in many-flavor gauge theories -- a conformal bootstrap approach},''
\href{http://arxiv.org/abs/1603.01995}{{\ttfamily arXiv:1603.01995 [hep-th]}}.
%%CITATION = ARXIV:1603.01995;%%.

\bibitem{Lipa:2003zz}
J.~Lipa, J.~Nissen, D.~Stricker, D.~Swanson, and T.~Chui, ``{Specific heat of
  liquid helium in zero gravity very near the lambda point},''
\href{http://dx.doi.org/10.1103/PhysRevB.68.174518}{{\em Phys.Rev.} {\bfseries
  B68} (2003) 174518}.
%%CITATION = PHRVA,B68,174518;%%.

\bibitem{Campostrini:2006ms}
M.~Campostrini, M.~Hasenbusch, A.~Pelissetto, and E.~Vicari, ``{The Critical
  exponents of the superfluid transition in He-4},''
  \href{http://dx.doi.org/10.1103/PhysRevB.74.144506}{{\em Phys.Rev.}
  {\bfseries B74} (2006) 144506},
\href{http://arxiv.org/abs/cond-mat/0605083}{{\ttfamily arXiv:cond-mat/0605083
  [cond-mat]}}.
%%CITATION = COND-MAT/0605083;%%.

\bibitem{Hasenbusch:2011yya}
M.~Hasenbusch, ``{Finite size scaling study of lattice models in the
  three-dimensional Ising universality class},''
  \href{http://dx.doi.org/10.1103/PhysRevB.82.174433}{{\em Phys.Rev.}
  {\bfseries B82} (2010) 174433},
\href{http://arxiv.org/abs/1004.4486}{{\ttfamily arXiv:1004.4486 [cond-mat]}}.
%%CITATION = PHRVA,B82,174433;%%.

\bibitem{SlavaUnpublished}
{Rychkov, S.} {\it unpublished work}.

\bibitem{Campostrini:2002ky}
M.~Campostrini, M.~Hasenbusch, A.~Pelissetto, P.~Rossi, and E.~Vicari,
  ``{Critical exponents and equation of state of the three-dimensional
  Heisenberg universality class},''
  \href{http://dx.doi.org/10.1103/PhysRevB.65.144520}{{\em Phys.Rev.}
  {\bfseries B65} (2002) 144520},
\href{http://arxiv.org/abs/cond-mat/0110336}{{\ttfamily arXiv:cond-mat/0110336
  [cond-mat]}}.
%%CITATION = COND-MAT/0110336;%%.

\bibitem{Caselle:2015csa}
M.~Caselle, G.~Costagliola, and N.~Magnoli, ``{Numerical determination of the
  operator-product-expansion coefficients in the 3D Ising model from
  off-critical correlators},''
  \href{http://dx.doi.org/10.1103/PhysRevD.91.061901}{{\em Phys. Rev.}
  {\bfseries D91} no.~6, (2015) 061901},
\href{http://arxiv.org/abs/1501.04065}{{\ttfamily arXiv:1501.04065 [hep-th]}}.
%%CITATION = ARXIV:1501.04065;%%.

\bibitem{Costagliola:2015ier}
G.~Costagliola, ``{OPE Coefficients of the 3D Ising model with a trapping
  potential},''
\href{http://arxiv.org/abs/1511.02921}{{\ttfamily arXiv:1511.02921 [hep-th]}}.
%%CITATION = ARXIV:1511.02921;%%.

\bibitem{DisorderedIsing}
Z.~Komargodski and D.~Simmons-Duffin, ``{The Random-Bond Ising Model in 2.01
  and 3 Dimensions},''
{\em to appear} (2015) .
%%CITATION = ARXIV:1511.02921;%%.

\bibitem{Beem:2013qxa}
C.~Beem, L.~Rastelli, and B.~C. van Rees, ``{The $\mathcal{N}=4$ Superconformal
  Bootstrap},'' \href{http://dx.doi.org/10.1103/PhysRevLett.111.071601}{{\em
  Phys.Rev.Lett.} {\bfseries 111} (2013) 071601},
\href{http://arxiv.org/abs/1304.1803}{{\ttfamily arXiv:1304.1803 [hep-th]}}.
%%CITATION = ARXIV:1304.1803;%%.

\bibitem{Beem:2015aoa}
C.~Beem, M.~Lemos, L.~Rastelli, and B.~C. van Rees, ``{The (2, 0)
  superconformal bootstrap},''
  \href{http://dx.doi.org/10.1103/PhysRevD.93.025016}{{\em Phys. Rev.}
  {\bfseries D93} no.~2, (2016) 025016},
\href{http://arxiv.org/abs/1507.05637}{{\ttfamily arXiv:1507.05637 [hep-th]}}.
%%CITATION = ARXIV:1507.05637;%%.

\bibitem{Poland:2011ey}
D.~Poland, D.~Simmons-Duffin, and A.~Vichi, ``{Carving Out the Space of 4D
  CFTs},'' \href{http://dx.doi.org/10.1007/JHEP05(2012)110}{{\em JHEP}
  {\bfseries 1205} (2012) 110},
\href{http://arxiv.org/abs/1109.5176}{{\ttfamily arXiv:1109.5176 [hep-th]}}.
%%CITATION = ARXIV:1109.5176;%%.

\bibitem{Poland:2015mta}
D.~Poland and A.~Stergiou, ``{Exploring the Minimal 4D $\mathcal{N}=1$ SCFT},''
  \href{http://dx.doi.org/10.1007/JHEP12(2015)121}{{\em JHEP} {\bfseries 12}
  (2015) 121},
\href{http://arxiv.org/abs/1509.06368}{{\ttfamily arXiv:1509.06368 [hep-th]}}.
%%CITATION = ARXIV:1509.06368;%%.

\bibitem{Iliesiu:2015qra}
L.~Iliesiu, F.~Kos, D.~Poland, S.~S. Pufu, D.~Simmons-Duffin, and R.~Yacoby,
  ``{Bootstrapping 3D Fermions},''
\href{http://arxiv.org/abs/1508.00012}{{\ttfamily arXiv:1508.00012 [hep-th]}}.
%%CITATION = ARXIV:1508.00012;%%.

\end{thebibliography}\endgroup
\bibliographystyle{utphys}

\end{document}